%% file: sample-sigconf.tex
\documentclass[sigconf, nonacm=true]{acmart}

\usepackage{booktabs} 
\usepackage{todonotes}
\usepackage{algorithm}
\usepackage{algpseudocode}
\usepackage{setspace}
\usepackage{subcaption}

\setcopyright{none}

\acmConference[Pre-print]{Pre-print}
\acmYear{2018}
\copyrightyear{2018}

\begin{document}
\title{Online Controlled Experiments for Personalised e-Commerce Strategies}
\subtitle{Design, Challenges, and Pitfalls}

\author{C. H. Bryan Liu}
\orcid{0000-0002-6516-2364}
\affiliation{%
  \institution{ASOS.com}
  \city{London} 
  \country{United Kingdom}
}
\email{bryan.liu@asos.com}

\author{Benjamin Paul Chamberlain}
\affiliation{%
  \institution{Imperial College London}
  \city{London} 
  \country{United Kingdom}
}
\email{}

\renewcommand{\shortauthors}{C. H. B. Liu et al.}

\begin{abstract}
Online controlled experiments are the primary tool for measuring the causal impact of product changes in digital businesses. It is increasingly common for digital products and services to interact with customers in a personalised way. Using online controlled experiments to optimise personalised interaction strategies is challenging because the usual assumption of statistically equivalent user groups is violated. Additionally,  challenges are introduced by users qualifying for strategies based on dynamic, stochastic attributes. Traditional A/B tests can salvage statistical equivalence by pre-allocating users to control and exposed groups, but this dilutes the experimental metrics and reduces the test power. 
We present a \emph{stacked incrementality test} framework that addresses problems with running online experiments for personalised user strategies. We derive bounds that show that our framework is superior to the best simple A/B test given enough users and that this condition is easily met for large scale online experiments.
In addition, we provide a test power calculator and describe a selection of pitfalls and lessons learnt from our experience using it.
\end{abstract}

%
%



\maketitle

\input{samplebody-conf}

\bibliographystyle{ACM-Reference-Format}
\bibliography{testing-bibliography-short} 

\end{document}

%% file: samplebody-conf.tex
\section{Introduction}

Online controlled experiments, which include A/B and multivariate tests, have become extremely popular over the past decade. Major technology companies such as Amazon~\cite{hill17efficient}, Airbnb~\cite{moss14experiment}, eBay~\cite{sadler15whynot}, Facebook~\cite{backstrom11network}, Netflix~\cite{xie16improving} and Yandex~\cite{poyarkov16boosted} have all reported the extensive use of online controlled experiments to measure the impact of their products and guide business decisions. Google~\cite{hohnhold15focusing}, Microsoft~\cite{kohavi13online}, and Linkedin~\cite{xu15frominfrastructure} report running hundreds or thousands of experiments concurrently on any given day. A number of start-ups (e.g. Optimizely~\cite{johari17peeking} and Qubit~\cite{browne17whatworks}) have recently been established purely to manage online controlled experiments for businesses.

\begin{figure}
\begin{center}
    \includegraphics[width=0.43\textwidth]{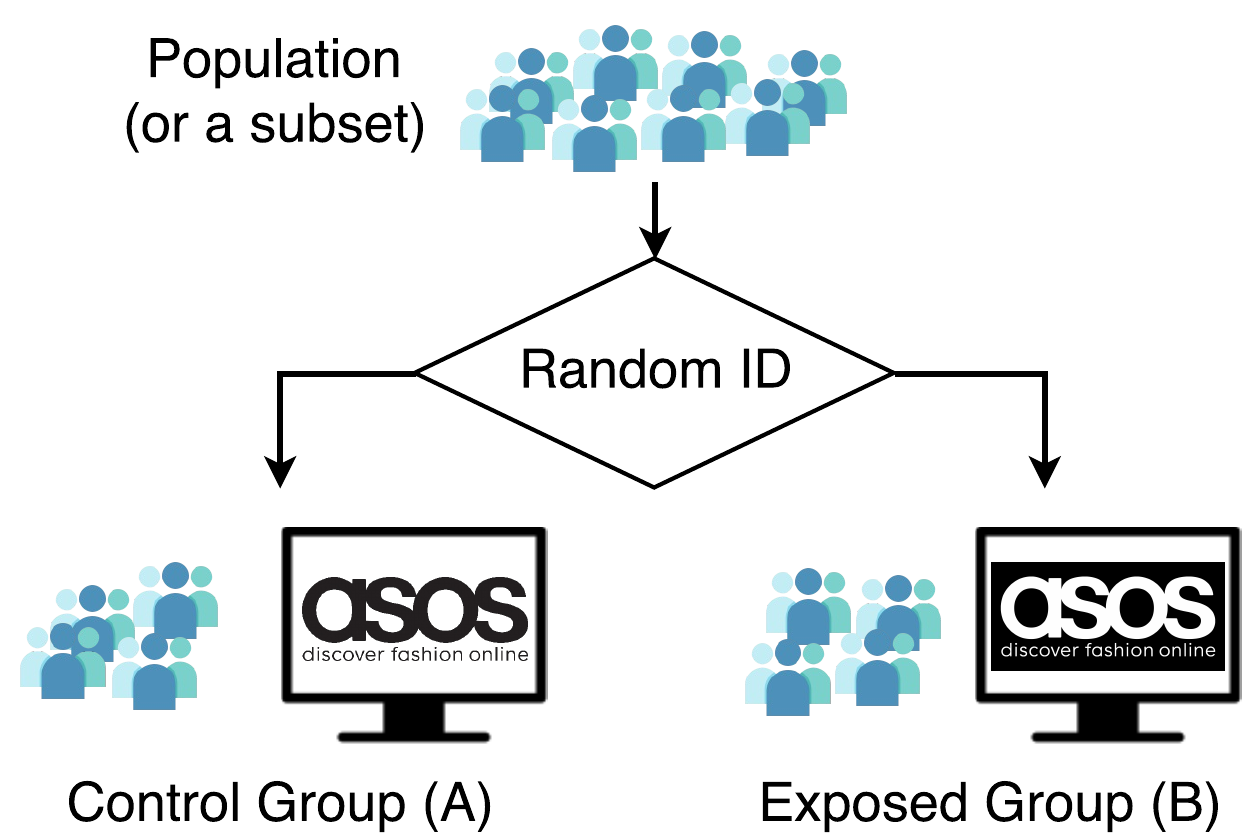}
\end{center}
\caption{Illustration of an A/B test set up. The test population is randomly split into a control group ($A$) and an exposed group ($B$). The control group is shown the existing logo, and the exposed group is exposed to a new logo. Metric(s) of interest are computed for both groups and compared against each other.}
\label{fig:AB_test_illustration_asos}
\end{figure}

The simplest form of online controlled experiments randomly splits individuals into two groups, $A$ and $B$. Figure \ref{fig:AB_test_illustration_asos} illustrates an A/B test for a new variant of the ASOS logo. Group $A$ is the control group and individuals allocated to this group see the existing logo. Group $B$ is the \textit{exposed} or \textit{treatment} group and individuals assigned to $B$ see a new variant of the logo. A business metric is measured for all individuals in the test and if the exposed group generates a statistically significant improvement in the metric, then the logo will be changed to the new variant.  Otherwise, the existing logo, often called the \textit{Business As Usual} (BAU) variant will remain. We call the difference in the business metric between the experimental variant and the BAU, the \textit{incrementality} of the variant.

In e-commerce it is often the case that instead of varying elements of the customer experience based on random splits, we want to experiment with variants that are personalised to individuals. In addition, instead of simple changes (like showing different logos) we want to compare complex sets of customer interactions, which we call \textit{strategies}. An example of a customer strategy is the scheduling, budgeting and ordering of marketing activities directed at an individual based on their purchase history. Running A/B tests to compare personalised customer strategies is significantly more complex than comparing simple product changes. As a running example, we use an experiment to measure the effect on customer purchase rate of a pop-up chat interface that appears after a customer has viewed either three of four products in a session. The challenges of running A/B tests to compare personalised customer strategies include:

\paragraph{Groups are not statistically equivalent}
Customers assigned to each strategy are not directly comparable because users cannot be randomly allocated to either strategy~1~or~2. Instead they must \textit{qualify} for a strategy through some action that is expected to correlate with the metric under measurement. In our example, customers who view four products are more likely to make a purchase. 
As the groups under comparison are not statistically equivalent it is not correct to simply choose the strategy that delivers the largest metric improvement. In our example, more customers will view three products in a session than four products and so delaying the pop-up until four products have been viewed may not be the best strategy, even if it is more incremental. Instead we must choose the strategy $x$ that delivers the most \textit{net benefit} (and use this as our test evaluation metric) defined as:
\begin{align}
\text{net benefit} = \text{incrementality} \times n_x ,
\end{align}
where $n_x$ is the projected number of customers who would receive strategy $x$.

\paragraph{Groups can not be allocated a priori}
Qualification for a strategy is dependent on dynamic, stochastic user attributes. We can not know at the start of a customer session how many products will be viewed or if a customer viewing three products will go on to view four. The distribution of users qualifying for each strategy can be estimated from data, but individual users qualify for each strategy stochastically.

\begin{figure}
\begin{center}
    \includegraphics[width=0.43\textwidth]{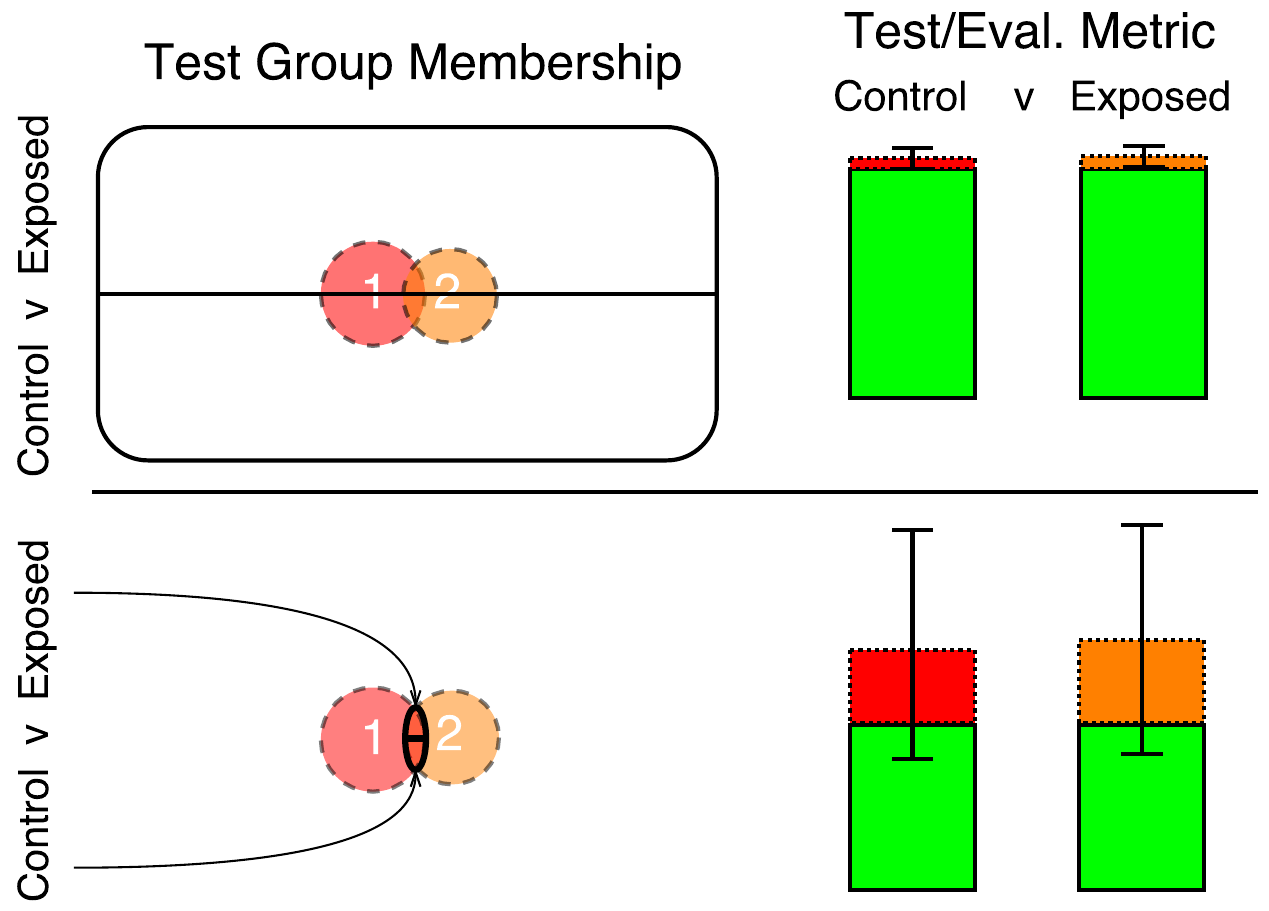}
\end{center}
\caption{
Numbered circles represent customers qualifying for different strategies. The bars on the right indicate the mean baseline metric (green) and incrementality from strategy~1 (red) and 2 (orange) with error bars. The top chart shows control/exposed groups sampled uniformly from the population (those in the rounded box), the effect on the metric is small due to dilution. The bottom chart show that limiting the control/exposed groups to those qualified for both strategies leads to few samples, resulting in high metric variance. 
}
\label{fig:test_group_membership_metric}
\end{figure}

\paragraph{Diluted metric movement}
Strategies are often only applicable to a small fraction of the total customers. If the control/exposed groups contain the entire customer population, any change in the mean of a metric will be heavily diluted by customers that did not qualify for the strategy (see Figure~\ref{fig:test_group_membership_metric}). Using our example, any effects on total purchase rate from customers who view three or more products will be diluted by customers who only view one or two products.
This effect is also observed by Lu and Liu~\cite{lu14separation}.

\paragraph{Lack of sample size/statistical power}
The power of a test is defined as the probability that the null hypothesis will be rejected given that the alternative hypothesis is true. To address the issue of dilution, it is possible to only compare customers who simultaneously qualify for both strategies $A$ and $B$. However, restricting the test subjects to this subset may result in insufficient test samples and hence underpowered tests, which is also illustrated in Figure~\ref{fig:test_group_membership_metric}. This problem is most acute when no customers simultaneously qualify for both strategies, as is the case in our running example. 


To address these challenges, we propose a \emph{stacked incrementality test} framework for optimising personalised user strategies. The framework handles groups that are not statistically equivalent and is more powerful than the best standard A/B test provided that sufficient numbers of users are available. We show that in practice, for large online experiments, this usually true. To the best of our knowledge, no test power calculator exists that can be adapted to our framework and so we provide an open source calculator for use by the community.

Our contributions are:
\begin{enumerate}
    \item We present a stacked incrementality framework for comparing different customer strategies making our code available on GitHub for the benefit of the wider community.\footnote{The code is available at \url{https://github.com/liuchbryan/stacked_incrementality_testing}}
    \item We derive bounds that express when the framework is superior to simpler experimental setups, show this usually to be true and provide a test power calculator.
    \item We describe a selection of pitfalls and lessons learnt from our experience of using the stacked incrementality test
\end{enumerate}

The rest of the paper is organised as follows. In Section~\ref{sec:existing_work} we summarise existing work in online controlled experiments. We present our stacked incrementality test framework in Section~\ref{sec:stacked_incrementality_test}, with a comparison to standard A/B tests following in Section~\ref{sec:evaluation} and lessons learnt using the framework in Section~\ref{sec:pitfalls_lessons}.

\section{Related Work}
\label{sec:existing_work}
A number of papers provide detailed advice on setting up, running, reporting and scaling online controlled experiments. The canonical example is the excellent work by~\citet{kohavi09controlled}, where the basic principles of good online experimental design are laid out. Several papers describe the challenges of building scalable, concurrent testing infrastructure at major technology companies \citep{kohavi13online,xu15frominfrastructure,tang10overlapping}.   \citet{kohavi14sevenrules} provide seven useful heuristics for running online experiments that go beyond technical instruction to include managing expectations of improvements from variants, handling results that are \textit{too good} and a version of Occam's Razor for online experimentation. \citet{xu16evaluating} adapt the general analysis of online testing to the specific case of mobile applications. In mobile applications some control of the test is surrendered to the app store and the users (who decide when to update the app) and they describe how to adapt to these constraints. 

One branch of research focusses on metric design and interpretation. \citet{deng16datadriven} discuss the need for meaningful metrics for online controlled experiments, and provided advice on how to choose, develop and evaluate the qualities of metrics. \citet{hohnhold15focusing} argue that it is more important to focus on the long-term effect of changes, and propose a method to estimate long-term effects from short-term metrics. A number of works from Microsoft~\cite{crook09sevenpitfalls,kohavi11unexpected,kohavi12trustworthy,dmitriev17adirtydozen} emphasise the need to properly understand \emph{why} metrics move and avoid various pitfalls while interpreting the metric movements.

Most work on online controlled experiments deals with observations that are approximately i.i.d.. In some cases this is a poor assumption such as when a recommender system is used to suggest products to users. In that case, the products create couplings between users and \citet{bakshy13uncertainty} have shown that this dependency results in an increased false positive rate when applied to Facebook data. To mitigate the problem, \citet{deng17trustworthy} proposed a variance estimation technique that accommodates a wide range of randomisation mechanisms in practical settings which does not necessarily require i.i.d. observations.

\citet{kohavi14sevenrules} note that ordinarily, successful experiments in technology companies only improve metrics by a fraction of a percent and \citet{xie16improving} describe the need to detect small effects as huge customer bases often translate them into substantial gains in revenue and profit. These observations motivate research into increasing the sensitivity of online tests by reducing measurement noise. Methods rely on decomposing the variance in a metric and attempting to eliminate unnecessary components. CUPED is a control variates method that measures a metric that is modified by a linear function of the covariates~\citep{deng13improving}. The idea is to eliminate the effect of correlated covariates on the the metric under measurement. A similar idea is used by \citet{poyarkov16boosted} who subtract the predicted value of a metric using boosted decision trees as predictors. In the same spirit, stratified sampling is used to eliminate variance between strata. Stratified sampling can be applied pre or post experiment and while the theoretical bounds are better for pre-experiment stratification, post-stratification is often preferable for large scale online A/B tests~\cite{xie16improving}. 


There is also research that specifically focuses on the problem of display ad targeting. \citet{hill15measuring} use causal inference as an alternative to A/B split testing to investigate the affect of an ad targeting algorithm by exploiting mediator variables. The mediator variable is whether the ad could be seen by the user as they found that only 50\% of ads served were seen by users.

\section{Stacked Incrementality Test}
\label{sec:stacked_incrementality_test}

Here we present the details of the \emph{stacked incrementality test} framework. We begin with a general description of how the test splits individuals into different groups and prescribes interventions. We then provide the mathematical detail of the framework, beginning with the assumptions and nomenclatures applied in Section \ref{sec:assumptions_nomenclatures}, followed by the derivation of the test statistic (Section \ref{sec:test_statistic}), test power (Section \ref{sec:test_power}), and minimum sample size required to run sufficiently powerful tests (Section \ref{sec:min_sample_size}). These calculations are necessarily for practitioners to design tests and conduct both \textit{a priori} and \textit{post-hoc} analyses under this framework.

\begin{figure}
\begin{center}
    \includegraphics[width=0.42\textwidth]{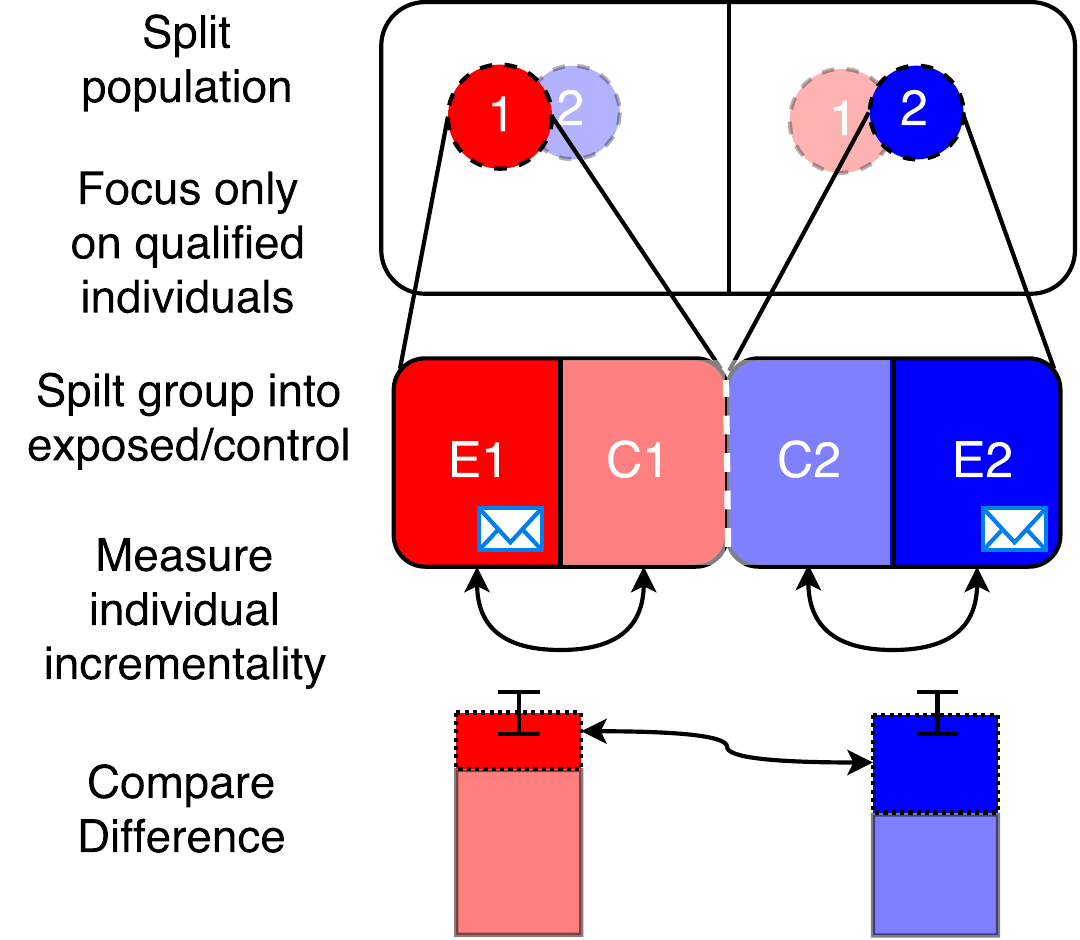}
\end{center}
\caption{Illustration of the setup of a stacked incrementality test, which a detailed description can be found in the beginning of Section~\ref{sec:stacked_incrementality_test}.
}
\label{fig:stacked_incrementality_test_design}
\end{figure}

The general design of the stacked incrementality test is illustrated in Figure \ref{fig:stacked_incrementality_test_design}. The population is randomly split into two groups. Any individual in the first group who qualifies for strategy~1, is then randomly allocated into a test ($E1$) or control ($C1$) group. In terms of our running example, customers in the first group are allocated to either $E1$ or $C1$ after viewing a third product. Those in $E1$ will then be shown a pop-up while those in $C1$ will never see a pop-up. Similarly, users in the second group qualify for strategy~2 when they view a fourth product and are split into $E2$ and $C2$. The second split is necessary as the requirement that customers qualify for a strategy breaks the statistical equivalence of the two groups.



By comparing group $E1$ to group $C1$, and group $E2$ to group $C2$ we can measure the incrementality of strategies~1~and~2 respectively. Once the incrementalities are established we 
compare if they are statistically different (see sections below).


In some cases $C1$ and $C2$ can be merged into a single control $C$. In the pop-up experiment, for instance, every individual in $C2$ can also be included in $C1$. 
Merging the control groups is desirable as the number of customers in each group can be increased, improving the test power. When this is the case the control groups $C1$ and $C2$ can be constructed from $C$ after the experiment based on individual attributes, with some individuals potentially included in both $C1$ and $C2$.


\subsection{Assumptions \& Nomenclatures}
\label{sec:assumptions_nomenclatures}
Here we introduce the formal notation necessary to evaluate the stacked incrementality framework.
Let $({R_g})_i$ be the \emph{response} of individual $i$ in group $g \in G = \{E1, E2, C1, C2\}$, which can be any business metric.

We begin by letting $({R_g})_i$ be an i.i.d. response within the same group $g$, with finite mean $\mu_g$ and (population) variance $\sigma^2_g$. The groups' mean and variance, while unknown, can be different. The i.i.d. assumption results in a more straightforward mathematical treatment.\footnote{For non-i.i.d. responses, we can estimate the variance using the technique proposed in \cite{deng17trustworthy}.}

The \emph{incrementality} in mean response (or net benefit, if used as a metric) within strategies~1~and~2 are respectively:
\begin{align}
    \mu_{D_1} = \mu_{E1} - \mu_{C1},\quad \mu_{D_2} = \mu_{E2} - \mu_{C2}, \label{eq:incrementality}
\end{align}

To select a strategy we must measure a statistically significant difference in incrementality
\begin{align}
    \mu_{D} = \mu_{D_2} - \mu_{D_1}.
    \label{eq:incrementality_CVR_across}
\end{align}
As we often test to see if a new strategy provides a better incrementality, we conduct a one-sided null hypothesis significance test on~$\mu_D$, where the null hypothesis is $\mu_D = 0$ and the alternate hypothesis being $\mu_D > 0$.\footnote{For two-sided test the mathematical treatment is similar.}





\subsection{Test Statistic}
\label{sec:test_statistic}
The test statistic is derived as follow.
Following Equation~\eqref{eq:incrementality} and~\eqref{eq:incrementality_CVR_across} the difference in sample means is given by
\begin{align}
    \overline{D} & \triangleq \overline{D}_2 - \overline{D}_1 = (\overline{R}_{E2} - \overline{R}_{C2}) - (\overline{R}_{E1} - \overline{R}_{C1}),
    \label{eq:sample_incrementality_mean} 
\end{align}
where $\overline{D}_1$ and $\overline{D}_2$ is the incrementality in sample mean response within strategies~1~and~2 respectively, and $\overline{R}_g$ is the sample mean response for group $g$. By the  Central Limit Theorem
\begin{align}
    \overline{R}_g \;\overset{\textrm{approx.}}{\sim}\; \mathcal{N}\left(\mu_g, \frac{\sigma^2_g}{n_g}\right),
    \label{eq:group_sample_normal_mean_distribution}
\end{align}
where $n_g$ denotes the number of individuals in group $g$. As $\overline{D}$ is a linear combination of $\overline{R}_{g}$: $g \in G$, $\overline{D}$ is distributed as
\begin{align}
    \overline{D} \;\overset{\textrm{approx.}}{\sim}\; \mathcal{N}\,\left(\mu_D, \sum_{g \in G}\frac{\sigma^2_g}{n_g}\right).
    \label{eq:sample_incrementality_approx_normal}
\end{align}


Welch's $t$-statistic in its general form, using the sample means $\overline{R}_g$ and sample variances $s^2_g$, follows approximately the Student's $t$-distribution $t_{\nu}$ with the degrees of freedom $\nu$ specified by the Welch-Satterthwaite Equation \cite{welch47,satterthwaite46}:
\begin{align}
    t = \frac{\overline{D}}{\sqrt{\sum_{g \in G}\psi_g}} \;\sim\; t_{\nu}, \quad \textrm{where}\;
    \nu = \frac{\left(\sum_{g \in G}\psi_g \right)^2}{\sum_{g \in G}\frac{\psi^2_g}{n_g - 1}} ,
    \label{eq:t_statistic_normal_population}
\end{align}

and $\psi_g = \frac{s^2_g}{n_g}$ is the estimated squared standard error of the sample mean $\overline{R}_g$. With a sufficiently large $n_g$ in every group $g$, the $t$-statistic in Equation~\eqref{eq:t_statistic_normal_population} effectively follows the standard normal distribution.\footnote{One might find while dealing with heavy-tailed distributions that the high sample variance precludes any meaningful conclusions being made. We recommend the use of variance reduction techniques mentioned in Section \ref{sec:existing_work} in this case.}

\subsection{Test Power}
\label{sec:test_power}
The power of a test is the probability that the difference in response incrementality is detected (i.e. there exist enough statistical evidence to reject the null hypothesis) when there is a difference in incremental effect (i.e. the alternate hypothesis is true)~\cite{ellis2010essential}.

We begin by observing the null hypothesis is rejected if $t > t_{\nu, 1-\alpha}$, the~$1-\alpha$ quantile for $t$-distribution with $\nu$~degrees of freedom. Under a \emph{specific} alternate hypothesis $\mu_D = \theta > 0$, the test power is specified as
\begin{align}
    1 - \beta_\theta & = \textrm{Pr}\,\Bigg(\frac{\overline{D}}{\sqrt{\sum_{g \in G} \psi_g}} > t_{\nu, 1-\alpha} \,\Bigg|\, \mu_D = \theta \Bigg),
    \label{eq:beta_theta_definition}
\end{align}
which we can show, similarly to the calculations in \cite{livingston05statistical,fleiss04determining}, is equivalent to
\begin{align}
    1 - \beta_\theta & = 1 - T_{\nu}\Bigg( t_{\nu, 1-\alpha} - \frac{\theta}{\sqrt{\sum_{g \in G} \psi_g }}\Bigg),
\end{align}
where $T_{\nu}$ denotes the Cumulative Density Function (CDF) of the Student's $t$-distribution with $\nu$ degrees of freedom.\footnote{Detailed steps of the calculation can be found on \url{https://github.com/liuchbryan/stacked_incrementality_testing}.}

To achieve a (pre-specified) minimum test power $\pi_{\textrm{min}} \in [0, 1]$, we require that
\begin{align}
    1 - \beta_\theta > \pi_{\textrm{min}}
    \iff 1 - \pi_{\textrm{min}} > T_{\nu}\Bigg( t_{\nu, 1-\alpha} - \frac{\theta}{\sqrt{\sum_{g \in G} \psi_g }}\Bigg) .
    \label{eq:beta_theta_gt_beta_star_ineq}
\end{align}
Assuming $\alpha < \pi_{\textrm{min}}$, which implies that $t_{\nu, 1-\alpha} > t_{\nu, 1-\pi_{\textrm{min}}}$, and taking the inverse CDF of the $t$-distribution (which is the quantile function), of both sides of Inequality~\eqref{eq:beta_theta_gt_beta_star_ineq} yields
\begin{align}
    & t_{\nu, 1-\pi_{\textrm{min}}} > t_{\nu, 1-\alpha} - \frac{\theta}{\sqrt{\sum_{g \in G} \psi_g }} 
    \label{eq:t_quantile_smaller_than_lhs_diff_ineq}\\
    \iff & \left(\frac{\theta}{t_{\nu, 1-\alpha} - t_{\nu, 1-\pi_{\textrm{min}}}}\right)^2 > \sum_{g \in G} \psi_g = \sum_{g \in G} \frac{s_g^2}{n_g} .
    \label{eq:square_smaller_than_sum_ineq}
\end{align}

In a \textit{post-hoc} analysis, a useful metric is the minimum effect the test will be able to detect under a test with power of at least $\pi_{\min}$. This can be found by recalling that $\theta$ is the effect specified by the specific alternate hypothesis. Rearranging Inequality \eqref{eq:t_quantile_smaller_than_lhs_diff_ineq} and fixing all $s^2_g$ and $n_g$, we observe:
\begin{align}
\theta > (t_{\nu, 1-\alpha} - t_{\nu, 1-\pi_{\textrm{min}}}) \sqrt{\sum_{g \in G} \psi_g}.
\label{eq:min_detectable_effect}
\end{align} 
Hence the minimum effect the test will be able to detect is given by the RHS term of Inequality \eqref{eq:min_detectable_effect}.

\subsection{Minimum Sample Size Required}
\label{sec:min_sample_size}
The test power is heavily dependent on the sample size of all groups, hence to achieve sufficient test power we often need to know \textit{a priori} the minimum number of samples required. In the simplest case all groups are of size $n$ and the test power can be determined by $n$. We also analyse the more complex case where group sizes differ in fixed ratios.

\subsubsection{Equal Sample Size In All Groups}

Our first scenario assumes that the sample size in all groups is equal, i.e. $n_g = n_{\textrm{min}} \,\forall g \in G$. Inequality \eqref{eq:square_smaller_than_sum_ineq} then becomes:
\begin{align}
     & \left(\frac{\theta}{t_{\nu, 1-\alpha} - t_{\nu, 1-\pi_{\textrm{min}}}}\right)^2 > \sum_{g \in G} \frac{s_g^2}{n_{\textrm{min}}} \\
    \iff & n_{\textrm{min}}  > \left(\frac{t_{\nu, 1-\alpha} - t_{\nu, 1-\pi_{\textrm{min}}}}{\theta}\right)^2 \sum_{g \in G} s_g^2 .
    \label{eq:n_min_gt_rhs}
\end{align}

Note we intentionally formulate Inequality \eqref{eq:n_min_gt_rhs} in such a way that sample size calculation can be generalised to as many groups with different estimated / measured sample variance as required.\footnote{The inequality assumes that the overall test statistic is a linear combination of each group's test statistic with weighting $\{+1, -1\}$. To apply a non unity weighting $k_g$ to group $g$, simply replace the $s^2_g$ term in Inequality \eqref{eq:n_min_gt_rhs} with $k^2_g s^2_g$.}

\subsubsection{Fixed Sample Size Ratio Between Groups}

The second scenario assumes that the groups are different sizes, but their relative sizes are fixed. This is often the case as the control group is typically smaller than the exposed group to limit the cost of withholding an action.
In the stacked incrementality test framework, using equal sized groups with a constant ratio for the control group, we put $n_{\min}$ individuals in each control group ($G_1 = \{C1, C2\}$), and $n_g$ individuals in each exposed groups ($G_2 = \{E1, E2\}$) such that $n_g = \frac{k_2}{k_1} n_{\min}$. 
Inequality \eqref{eq:square_smaller_than_sum_ineq} then becomes
\begin{align}
    & \left(\frac{\theta}{t_{\nu, 1-\alpha} - t_{\nu, 1-\pi_{\textrm{min}}}}\right)^2 > \sum_{g \in G_1} \frac{s_g^2}{n_{\min}} + \sum_{g \in G_2} \frac{s_g^2}{n_g} \\
    \iff & n_{\min} > \left(\frac{t_{\nu, 1-\alpha} - t_{\nu, 1-\pi_{\textrm{min}}}}{\theta}\right)^2 \left(\sum_{g \in G_1} s_g^2 + \sum_{g \in G_2} \frac{k_1s_g^2}{k_2} \right) .
\end{align}
This can be easily extended to cover cases with multiple different sized partitions, providing the ratio between each partition is specified.

\section{Comparison with A/B Tests}
\label{sec:evaluation}

\begin{figure}
\begin{center}
    \includegraphics[width=0.47\textwidth, trim = 0 1mm 28mm 0, clip]{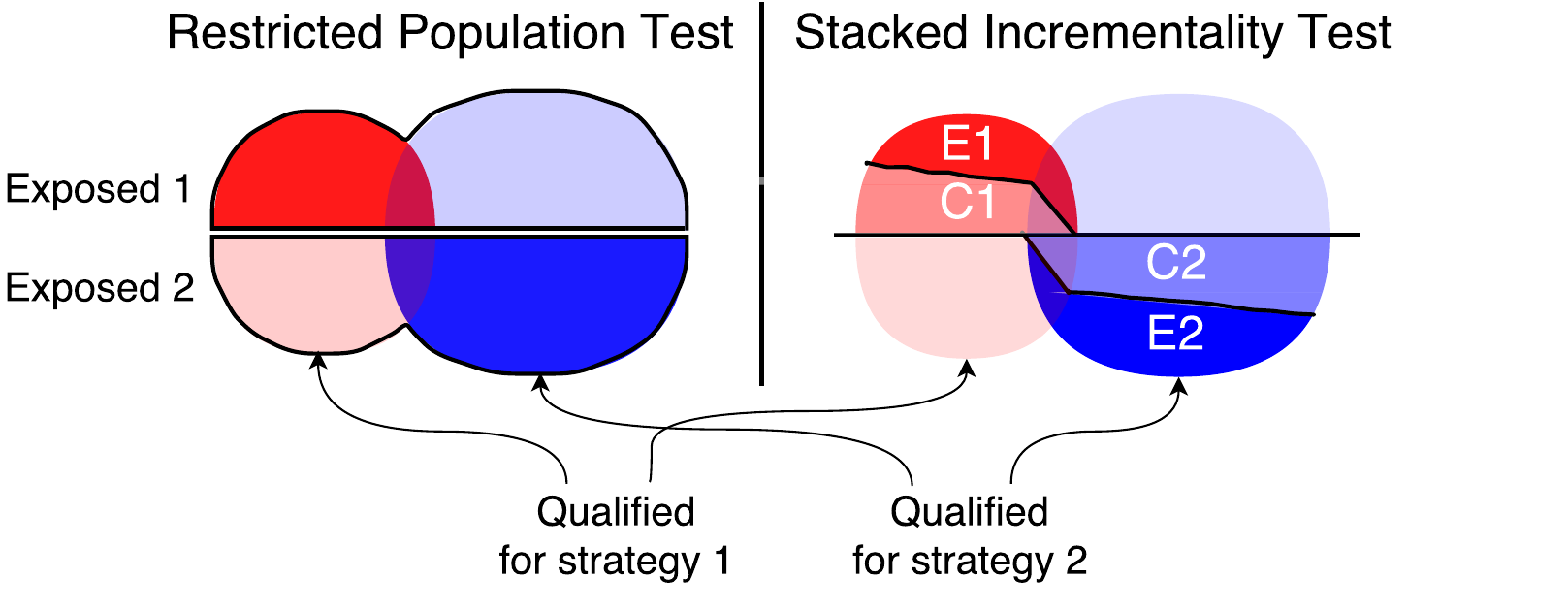}
\end{center}
\caption{Group composition in the Restricted Population A/B Test (RPT) and Stacked Incrementality Test (SIT). The left red (right blue) circle under each test represents those who qualifies for strategy~1 (2). The solid colour represents those who receive an intervention under the respective strategies, and those in lighter colour(s) receive the control action. See the beginning of Section \ref{sec:evaluation} for a detailed description of RPT.}
\label{fig:comparison_segments}
\end{figure}

Here we compare the Stacked Incrementality Test (SIT) to a standard A/B test that contains the minimum number of individuals required to ensure statistical equivalence. Figure \ref{fig:comparison_segments} shows the individuals included in this Restricted Population Test (RPT) and those in the SIT. The RPT includes only individuals who qualify for either strategy. Figure \ref{fig:comparison_segments} shows that this subset is split into those that could be exposed to strategy~1 (above the line) and those that could be exposed to strategy~2 (below the line). The bold fill colour indicates individuals who received an intervention ie. both qualified for a strategy and were randomly allocated to that strategy's group.

In this section we derive the incrementality difference that can be detected by both RPT and SIT given the distributional parameters of the test groups (and their subsets). Then we define the minimum detectable effect for both tests. We use these results to formulate conditions under which SIT is superior to RPT and show that this is true given mild assumptions that are usually true of large scale online experiments. 

\begin{figure}
\begin{center}
\includegraphics[width=0.47\textwidth, trim = 0 2mm 0 0, clip]{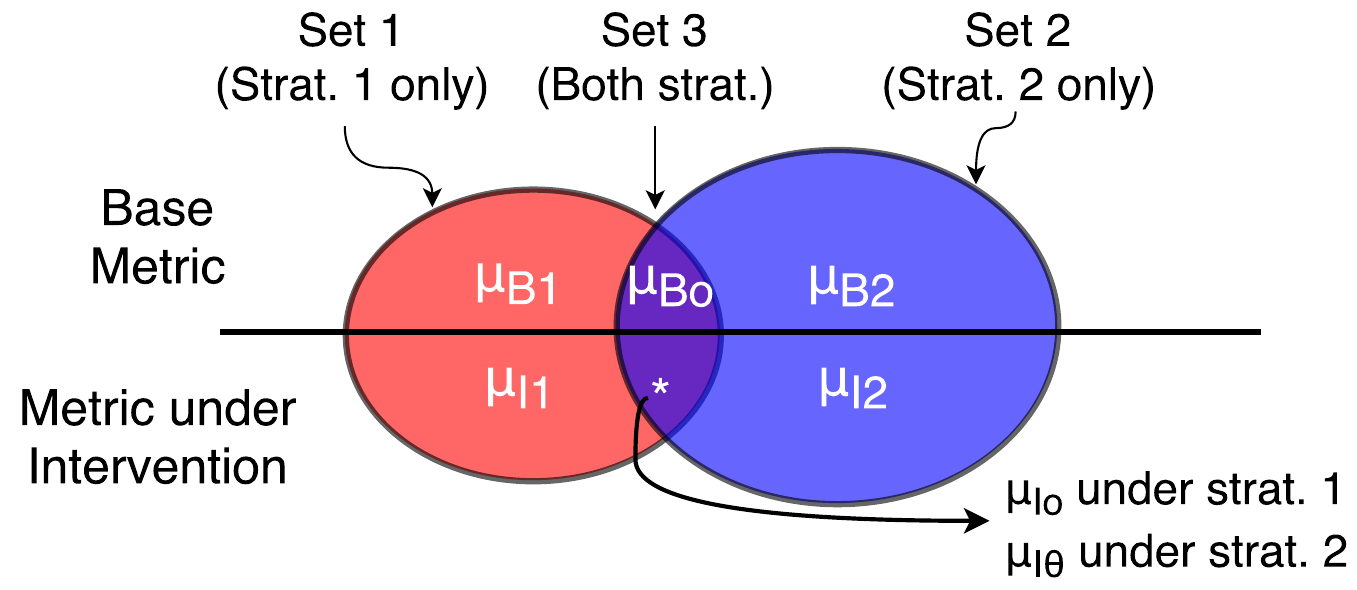}
\end{center}
\caption{The means of the base metric and the metric under intervention for those who qualify for strategy~1 only (Set 1), qualify for strategy 2 only (Set 2), and qualify for both strategies (Set 3). 
}
\label{fig:comparison_metric_assumption}
\end{figure}

The individuals in both tests can be decomposed into three groups: (1) those who qualify for strategy~1 only, (2) those who qualify for strategy 2 only, and (3) those who qualify for both strategies. Figure \ref{fig:comparison_metric_assumption} shows this decomposition and the associated mean metric values for each subset. 
Let $(\mu_{B1}, \sigma^2_{B1})$, $(\mu_{B2}, \sigma^2_{B2})$, and $(\mu_{Bo}, \sigma^2_{Bo})$ be the mean and variance for the base metric value of sets (1), (2), and (3) respectively.\footnote{The last pair is necessary as we cannot assume those who qualify for both strategies have their base metric value as a linear combination of those who only qualify for one strategy only.} For those in sets (1) and (2), let $(\mu_{I1}, \sigma^2_{I1})$, and $(\mu_{I2}, \sigma^2_{I2})$ be the mean and variance for the metric value under the intervention prescribed by the respective strategies. In addition let $(\mu_{Io}, \sigma^2_{Io})$, and $(\mu_{I\theta}, \sigma^2_{I\theta})$ be the mean and variance for the metric value of set (3) under the intervention prescribed by strategies 1 and 2 respectively.\footnote{There is no reason to assume that the subset which qualifies for both strategies will react the same way to both strategies. At the same time, no one in this set will be prescribed an intervention from both strategies.} Finally let $n_1$, $n_2$, and $n_o$ be the number of individuals in sets (1), (2), and (3) respectively.

We first look at the incrementality difference reported by RPT (denoted $\Delta_{R}$). The test has two groups, each taking roughly half of the three sets. For the second group (that below the line in Figure~\ref{fig:comparison_segments}), we prescribe interventions from strategy~2 on sets~(2) and~(3), leaving set~(1) receiving the control action, therefore the mean of the metric in this group is
\begin{align}
    \frac{n_2}{n_{U}}\mu_{I2} + 
    \frac{n_o}{n_{U}}\mu_{I\theta} + 
    \frac{n_1}{n_{U}}\mu_{B1} ,
\end{align}
where $n_U = n_1 + n_2 + n_o$ is the number of individuals in a RPT. Applying similar treatment to the first group (that above the line in Figure~\ref{fig:comparison_segments}) we obtain another mean. The incrementality difference is the difference in these two means: 
\begin{align}
    \Delta_{R} = \frac{1}{n_U}\big(&(n_2\mu_{I2} + n_o\mu_{I\theta} + n_1\mu_{B1}) \nonumber\\ 
    & - (n_2\mu_{B2} + n_o\mu_{Io} + n_1\mu_{I1})\big).
    \label{eq:delta_RPT_final}
\end{align}

Now we recall the SIT has four groups, each taking roughly one fourth of the sets required by the strategies (i.e. groups $E1$ and $C1$ taking those in sets (1) and (3), and groups $E2$ and $C2$ taking those in sets (2) and (3)). Take group $E2$ as an example, the expected mean of this group is the weighted mean of the corresponding sets after intervention:
\begin{align}
    \frac{n_2}{n_2+ n_o} \mu_{I2} + 
    \frac{n_o}{n_2 + n_o} \mu_{I\theta} .
\end{align}
Applying the same reasoning to the other groups and taking the difference between the incrementality of strategies 1 and 2 leads to a stacked incrementality difference of
\begin{align}
    \Delta_{S} \triangleq &
    \big(\frac{n_2}{n_2 + n_o}\mu_{I2} + \frac{n_o}{n_2 + n_o}\mu_{I\theta}\big) 
    - \big(\frac{n_2}{n_2 + n_o}\mu_{B2} + \frac{n_o}{n_2 + n_o}\mu_{Bo}\big) \nonumber\\
    & - \Big(\big(\frac{n_1}{n_1 + n_o}\mu_{I1} + \frac{n_o}{n_1 + n_o}\mu_{Io}\big)
    - \big(\frac{n_1}{n_1 + n_o}\mu_{B1} + \frac{n_o}{n_1 + n_o}\mu_{Bo}\big)\Big) .
\end{align}
Combining the fractions gives
\begin{align}
    \Delta_{S} = &
    \frac{1}{n_2 + n_o} [n_2(\mu_{I2} - \mu_{B2}) + n_o(\mu_{I\theta} - \mu_{Bo})] \nonumber\\
    & - \frac{1}{n_1 + n_o} [n_1(\mu_{I1} - \mu_{B1}) + n_o(\mu_{Io} - \mu_{Bo})].
    \label{eq:delta_s_final}
\end{align}

We then consider the minimum detectable effect provided by both tests.
From Equation \eqref{eq:min_detectable_effect}, the minimum detectable effect is given by:
\begin{align}
    \theta_{\min} = (z_{1-\alpha} - z_{1-\pi_{\min}}) \sqrt{\sum_{g \in G} \frac{\sigma^2_g}{n_g}},
\end{align}
where we use the population variances $\sigma^2$ instead of the sample variances $s^2$, and normal quantiles instead of the $t_\nu$ quantiles.\footnote{This is because there is no longer any variability in the variance, and hence the test statistic follows a normal distribution instead of a Student's $t$-distribution. In the case where we use the sample variances $s^2$ and perform more rigorous calculation, we have to be careful on the small, but often negligible, difference between the difference of $t_\nu$ quantiles under different tests, as they are likely to have a different degrees of freedom~$\nu$.} For RPT, the variance of the first group's mean metric value is
\begin{align}
    \frac{\frac{n_1}{2}\sigma^2_{I1} + \frac{n_o}{2}\sigma^2_{Io} + \frac{n_2}{2}\sigma^2_{B2}}{\left(\frac{n_U}{2}\right)^2},
\end{align}
which is the weighted variance of all sets, divided by the number of individuals in the group.
Applying the same treatment to the second group leads to a minimum detectable effect for RPT of
\begin{align}
 \theta_{R} \triangleq z \sqrt{\frac{2(n_1\sigma^2_{I1} + n_o\sigma^2_{Io} + n_2\sigma^2_{B2})}{\left(n_U\right)^2} + \frac{2(n_1\sigma^2_{B1} + n_o\sigma^2_{I\theta} + n_2\sigma^2_{I2})}{\left(n_U\right)^2}} ,
\end{align}
where $z = z_{1-\alpha} - z_{1-\pi_{\min}}$ is the difference of the normal quantiles. Again, assuming $\alpha < \pi_{\min}$ and $z$ is positive. Pulling the common fraction out of the square root, and grouping the remaining terms, we arrive at
\begin{align}
 \theta_{R} = \frac{\sqrt{2}\,z}{n_U} \sqrt{n_1(\sigma^2_{I1} + \sigma^2_{B1}) + n_2(\sigma^2_{I2} + \sigma^2_{B2}) + n_o(\sigma^2_{Io} + \sigma^2_{I\theta}) }.
  \label{eq:theta_min_RPT}
\end{align}

To calculate the minimum detectable effect of SIT we begin by deriving the variance of group $E1$ among the four groups, which is the weighted variance between sets (1) and (2), divided by the number of individuals in the group:
\begin{align}
    \frac{\frac{n_1}{4}}{\left(\frac{n_1}{4} + \frac{n_o}{4}\right)^2}\sigma^2_{I1} + \frac{\frac{n_o}{4}}{\left(\frac{n_1}{4} + \frac{n_o}{4}\right)^2}\sigma^2_{Io}.
\end{align}
With a similar derivation for the variance of the other three groups. The denominators in each variance are the same and hence can be combined, leading to a minimum detectable effect for SIT of
\begin{align}
    \theta_S \triangleq z \sqrt{
    \begin{array}{l}
        \frac{4(n_1\sigma^2_{I1} + n_o\sigma^2_{Io})}{(n_1 + n_o)^2}
        + \frac{4(n_1\sigma^2_{B1} + n_o\sigma^2_{Bo})}{(n_1 + n_o)^2} \\[0.6em]
        + \frac{4(n_2\sigma^2_{B2} + n_o\sigma^2_{Bo})}{(n_2 + n_o)^2}
        + \frac{4(n_2\sigma^2_{I2} + n_o\sigma^2_{I\theta})}{(n_2 + n_o)^2}
    \end{array}
    }.
\label{eq:sit1}
\end{align}

To directly compare Equation~\eqref{eq:sit1} to Equation~\eqref{eq:theta_min_RPT} we pull a factor of $2/n_U$ out of the square root and collect $n$-terms leading to
\begin{align}
    \theta_S = \frac{2z}{n_U} \sqrt{
    \begin{array}{l}
        n_1\left(\frac{n_U}{n_1 + n_o}\right)^2 (\sigma^2_{I1} + \sigma^2_{B1})
        + n_2\left(\frac{n_U}{n_2 + n_o}\right)^2 (\sigma^2_{I2} + \sigma^2_{B2}) \\[0.6em]
        + n_o \left[\left(\frac{n_U}{n_1 + n_o}\right)^2 (\sigma^2_{Io} + \sigma^2_{Bo}) + \left(\frac{n_U}{n_2 + n_o}\right)^2 (\sigma^2_{I\theta} + \sigma^2_{Bo})
        \right]
    \end{array}
    }.
    \label{eq:theta_min_SIT}
\end{align}

By inspecting Equations \eqref{eq:theta_min_RPT} and \eqref{eq:theta_min_SIT}, $\theta_{S} > \theta_{R}$ and SIT will always be less sensitive than RPT. This is because $\theta_{S}$ features extra non-negative terms and multipliers greater or equal to one. Therefore SIT is superior to RPT when the gain in incrementality difference is greater than the loss in sensitivity:
\begin{align}
    \Delta_S - \Delta_{R} > \theta_S - \theta_{R} .
    \label{eq:SIT_superior_condition}
\end{align}

Here we analyse the condition for an important special case, where $n_o = 0$ (i.e. there is no overlap between the strategies' audiences) and $n_1 = n_2 = n$ (i.e. both strategies have the same number of individuals). The LHS of Inequality \eqref{eq:SIT_superior_condition} then becomes:
\begin{align}
    \Delta_S - \Delta_{R}  = &
    \frac{1}{n}\big[n(\mu_{I2} - \mu_{B2})\big] - \frac{1}{n}\big[n(\mu_{I1} - \mu_{B1})\big] \nonumber\\
    & - \frac{1}{n+n}\big[(n\mu_{I2} + n\mu_{B1}) - (n\mu_{B2} + n\mu_{I1})\big] .
\end{align}
Cancelling terms in $n$ and grouping on $\mu$ leads to
\begin{align}
    \Delta_S - \Delta_{R}  = &
    \frac{1}{2}[(\mu_{I2} - \mu_{B2}) - (\mu_{I1} - \mu_{B1})] \,,
    \label{eq:Delta_S_minus_Delta_U_final}
\end{align}
which depends only on the incrementality difference between strategies. For the RHS of Inequality  \eqref{eq:SIT_superior_condition}, we begin with:
\begin{align}
    \theta_S - \theta_{R} & = \nonumber\\
    \frac{z}{n+n}\Big[&2\sqrt{n \left(\frac{n+n}{n}\right)^2 (\sigma^2_{I1} + \sigma^2_{B1}) + n \left(\frac{n+n}{n}\right)^2 (\sigma^2_{I2} + \sigma^2_{B2})} \nonumber\\
         & - \sqrt{2}\sqrt{n(\sigma^2_{I1} + \sigma^2_{B1}) + n(\sigma^2_{I2} + \sigma^2_{B2})}\Big] \,.
\end{align}
Combining the terms within each square root we get:
\begin{align}
    \theta_S - \theta_{R} =
    \frac{z}{2n}\Big[&4\sqrt{n (\sigma^2_{I1} + \sigma^2_{B1} + \sigma^2_{I2} + \sigma^2_{B2})} \nonumber\\
         & - \sqrt{2}\sqrt{n(\sigma^2_{I1} + \sigma^2_{B1} + \sigma^2_{I2} + \sigma^2_{B2})}\Big] .
\end{align}
Combining the square root terms using the sum of squares, and pulling the preceding constant and $n$ out of the square root, we arrive at:
\begin{align}
    \theta_S - \theta_{R} =
    \frac{(2 - \frac{1}{\sqrt{2}})z}{\sqrt{n}} \sqrt{\sigma^2_{I1} + \sigma^2_{B1} + \sigma^2_{I2} + \sigma^2_{B2}} \,,
    \label{eq:theta_S_minus_theta_S_final}
\end{align}
which depends on both the variances of the sets and the size of the groups (recall here $n$ represents the number of people who qualify for strategy~1 (or 2) only, not the combined count).

Combining Inequality \eqref{eq:SIT_superior_condition} and Equations \eqref{eq:Delta_S_minus_Delta_U_final} and \eqref{eq:theta_S_minus_theta_S_final}, we conclude that SIT is superior to RPT if
\begin{align}
    \frac{1}{2}[(\mu_{I2}& - \mu_{B2}) - (\mu_{I1} - \mu_{B1})]  > \nonumber\\
    &\frac{(2 - \frac{1}{\sqrt{2}})z}{\sqrt{n}} \sqrt{\sigma^2_{I1} + \sigma^2_{B1} + \sigma^2_{I2} + \sigma^2_{B2}}
\end{align}
\begin{align}
    \iff n > 
    \left(\frac{(4 - \sqrt{2})\,z\, \sqrt{\sigma^2_{I1} + \sigma^2_{B1} + \sigma^2_{I2} + \sigma^2_{B2}}}
    {[(\mu_{I2} - \mu_{B2}) - (\mu_{I1} - \mu_{B1})]}\right)^2 ,
    \label{eq:SIT_superior_condition_in_n}
\end{align}
assuming, without loss of generality, that strategy~2 gives a larger incrementality than strategy~1. 

To analyse the inequality we consider an experiment to optimise conversion rate, which has well bounded values for $\sigma$ and $\mu$. In this case the variance of a sample conversion rate $p \in [0, 1]$ is given as $p\,(1-p)$, which attains its maximum value of a quarter when $p=0.5$. Hence the sum of four variances cannot exceed one. This leads to a lower bound where the minimum number of individuals required for a strategy is a function of the incrementality difference between two strategies~$\Delta$:
\begin{align}
    n > \frac{\left((4-\sqrt{2})\,z\right)^2}{\Delta^2} \approx \frac{41.34}{\Delta^2},
    \label{eq:strategy_size_strong_bound}
\end{align}
where $\Delta = (\mu_{I2} - \mu_{B2}) - (\mu_{I1} - \mu_{B1})$. Therefore, SIT is superior to RPT in detecting a 0.5\% absolute incrementality difference with 1.66M individuals in each strategy and a 2.5\% absolute difference with 67k individuals. These values are clearly easily within the scope of modern online controlled experiments, which may use tens or even hundreds of millions of users.

\section{Common Pitfalls \& Lessons Learnt}
\label{sec:pitfalls_lessons}
Having described the SIT framework and outlined the conditions under which it is superior to a naive A/B test, we share some useful lessons for practitioners who wish to use the SIT. While there are many papers dealing with common pitfalls in general A/B tests~\cite{crook09sevenpitfalls,kohavi11unexpected,kohavi12trustworthy,kohavi14sevenrules,xu15frominfrastructure,dmitriev17adirtydozen,deng17trustworthy}, here we describe only  experiments that consider two personalised customer \textit{strategies} where a strategy only applies to a small subset of customers and the customers in each strategy are not statistically equivalent.

\subsection{Cross-contamination Between Strategies}
\label{sec:test_cross_contamination}

We used the stacked incrementality framework as part of a strategy to reduce the churn rate at ASOS. The baseline strategy emailed customers a discount coupon if they had not engaged with the website for a period of time $t$.


We set up a stacked incrementality test to compare the existing strategy with a new strategy that emailed customers if their predicted churn risk \cite{chamberlain17customer} exceeded a threshold $c$. The goal of the experiment was to improve the incremental purchase rate of customers receiving the email.

\begin{figure}
\begin{center}
    \includegraphics[width=0.45\textwidth, trim = 0 0 5mm 0]{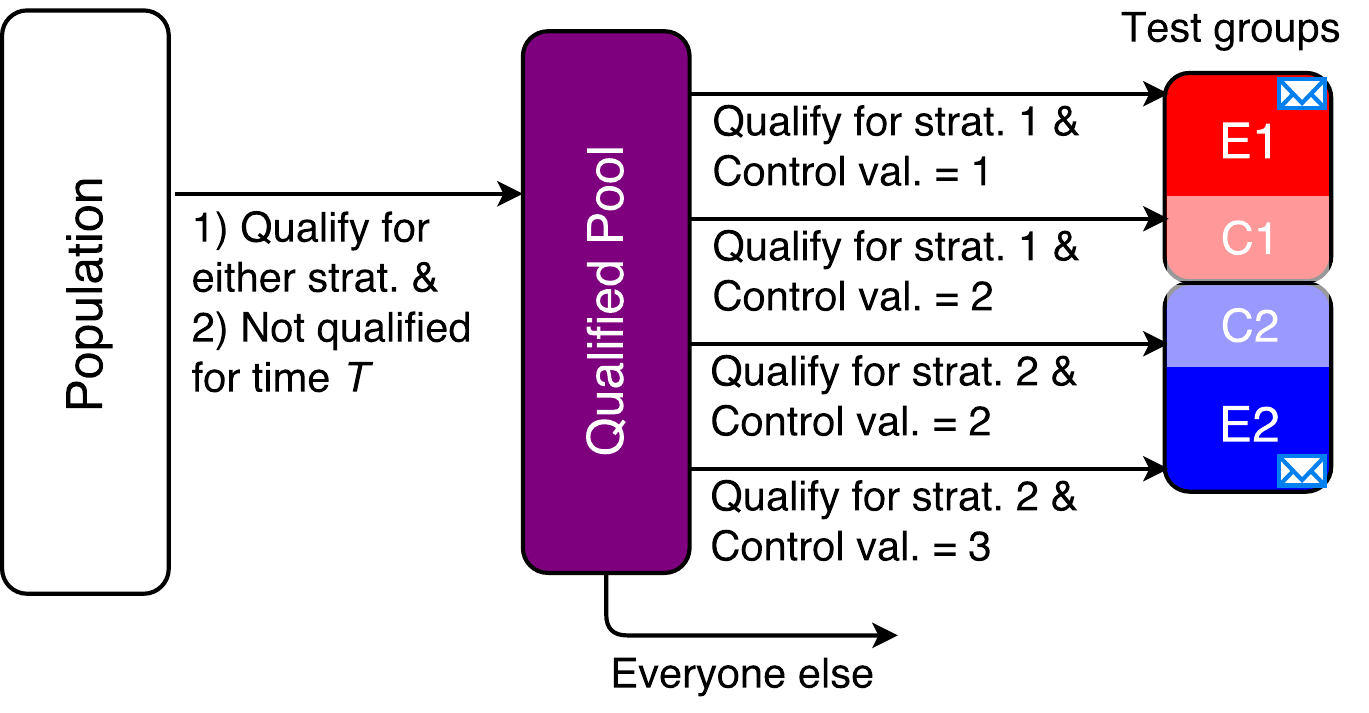}
\end{center}
\caption{The stacked incrementality test design for comparing email strategies. See Section \ref{sec:test_cross_contamination} for a detailed description of the setup.}
\label{fig:SIT_errorous_setup}
\end{figure}

The test design is shown in Figure \ref{fig:SIT_errorous_setup}. Customers who qualified for either strategy on any given day are added to the qualified pool, where they are split by a randomised control value (RCV). Those who have a RCV of one and have not engaged with the business for time $t$ (i.e. qualifying for strategy~1) are put in group $E1$ (exposed group of strategy~1). Those who have a RCV of three and have a churn risk greater than $c$ (i.e. qualifying to strategy~2) are put in group $E2$. Those who have a RCV of two are put into the control groups depending on which strategy they qualified for. Customers in the exposed groups are sent marketing emails with a discount code, and those in the control groups, together with those in the qualified pool who are not assigned to a group, are not emailed. Customers who qualified for either strategy within time $T > t$ were excluded to prevent multiple emails being sent on consecutive days.

\begin{figure}
    \begin{subfigure}[b]{0.42\textwidth}
        \begin{center}
            \includegraphics[width=\textwidth]{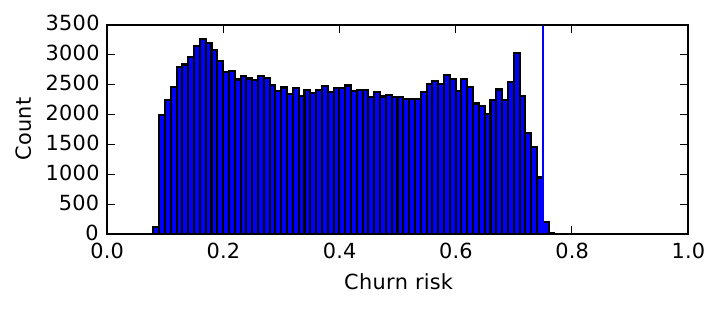}
        \end{center}
        \vspace*{-10pt}
        \caption{Exposed group of strategy~1 ($E1$)}
    \end{subfigure}
    \begin{subfigure}[b]{0.415\textwidth}
        \begin{center}
            \includegraphics[width=\textwidth]{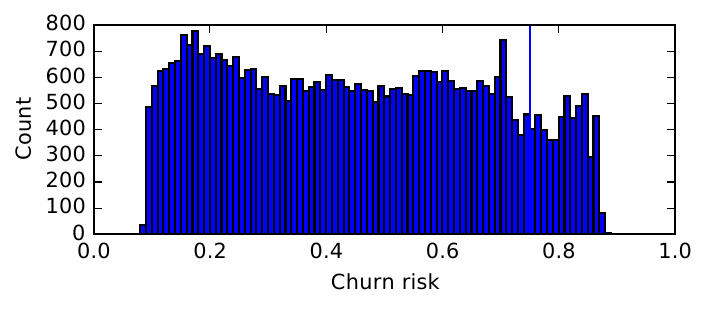}
        \end{center}
        \vspace*{-10pt}
        \caption{Control group of strategy~1 ($C1$)}
    \end{subfigure}
\caption{The distribution of churn risk among individuals qualified for strategy~1, which does not take into account the risk in disengaging. The distributions should be statistically equivalent but is not the case as observed, with those in the exposed group with a high risk (right of blue vertical line) excluded. Note strategy~1 only targets a small group of the entire customer base and hence the plots does not represents the overall churn risk distribution.}
\label{fig:risk_distribution}
\end{figure}

During the test, our monitoring system flagged anomalies in several data quality metrics. These included a mismatch in the distribution of the churn risk between groups~$E1$ and~$C1$, which should have been statistically equivalent (see Figure \ref{fig:risk_distribution}), as well as a sample ratio mismatch between the corresponding exposed and control groups.\footnote{See Dmitriev et al. \cite{dmitriev17adirtydozen} for more detail in data quality metrics and sample ratio mismatch.} Intriguingly, only customers with high churn risk were missing from $E1$, while the distributions at lower churn risks were similar across groups~$E1$ and~$C1$. We initially suspected that the system was first allocating all customers with a churn risk greater than~$c$ to~$E2$ and only those remaining were considered for~$E1$. However, on closer examination we discovered that excluding customers from re-entering the experiment within time $T$ was the source of the trouble. Consider a customer with a high churn risk and a RCV of one. Based on their RCV they are routed into group~$E1$ if they have not engaged for time~$t$. However, they are included into the qualified pool (as a strategy~2 target) due to their high churn risk. Once inside the qualified pool they are dropped because they have the ``wrong'' control value. Having ``qualified'' they are now prevented from being emailed for time~$T$, which is longer than the duration of the test. This phenomena caused the mismatch in distributions observed in Figure~\ref{fig:risk_distribution}. The control groups are not affected as they share the same RCV in this case.

We call this effect \emph{cross-contamination} between strategies as a result of an incorrect test setup. The subtle error leads to strategy~2 ``barring'' some individuals from qualifying for strategy~1, and to a lesser extent vice versa. The strategies are effectively meddling with each others audience composition, hence contaminating the metrics and rendering the test result unusable.

\subsection{Contamination Across Exposed / Control Groups}
\label{sec:test_group_contamination}


We ran a similar experiment to the one described in Section \ref{sec:test_cross_contamination} at the end of 2017 and found that not only was the incrementality difference insignificant, but we were unable to show that either of the strategies provided a significant uplift. We suspected that the test under powered, but upon further investigation found the existing strategy to have a much smaller effect than previous tests using the same numbers of samples. However, none of the data quality metrics indicated anything suspicious.

We only realise later that the test covered the lead up to and the duration of the Black Friday weekend. Black Friday is the largest sale event in the year, and an unusual distribution of customers are attracted to the site by  well-publicised and hefty-sized discounts. To make matters worse, these discounts were typically larger than those offered by our email, which struggled to stand out against the cacophony of marketing activity at that time. As a result, we observed, \textit{post-hoc}, 20 times more customers taking up the Black Friday sales discounts than the email discount in the exposed groups, and a similar number of customers taking up the Black Friday sales discount in the control groups during the test period. This led to the incrementality of the email being diluted compared to previous tests that were conducted outside of sales periods.

In contrast to Section \ref{sec:test_cross_contamination} where there is a cross-contamination between strategies, here we have metrics in all groups of the test contaminated by an external event. This happens when the external event has a greater influence on the metric than any interventions carried out under the experimental strategies, and can happen even when the test is set up correctly. 


\subsection{A Proxy for Control Groups That Went Wrong}
\label{sec:control_proxy}

We finally describe an experiment in programmatic marketing aimed at improving the bidding strategy for retargeting adverts. A retargeting advert displays products that customers have previously viewed, but not purchased, when they visit websites that sell advertising space. Advertising space is bought by bidding in online real-time auctions to show an ad to a single visitor. A bidding strategy refers to a particular allocation of marketing spend to auctions based on customer attributes.
In our experiment, the baseline strategy segmented individuals on a number of attributes including gender, use of desktop/mobile device and the time since their last session (recency). For the exposed group we replaced recency buckets with predictive customer lifetime value~\cite{chamberlain17customer} buckets. The metrics measured were the mean spend per customer and the conversion rate.

The test reported a large lift in conversion rate and mean spend per person targeted. 
As the size of the improvement was larger than expected, we applied Twyman's law (``Any figure that looks interesting or different is usually wrong.''~\cite{kohavi14sevenrules,kohavi13online,dmitriev17adirtydozen}) and began searching for potential errors in the test methodology. Our analysis revealed some disturbing results --- the original bidding strategy had a \emph{negative} effect on conversion rate (-10.8\% absolute) and mean spend per person targeted (-{\pounds}22.3 absolute). The new bidding strategy was also negative, though to a lesser extent, making it appear to be the superior in the stacked framework.


Again, the test setup was flawed. Any customer who viewed a product in a session qualified for retargeting, unless they made a purchase within the session. 
While it was easy to tell which customers in the exposed groups were served ads, it is very hard to know who in the control groups \textit{could} have been served an ad. As a proxy, we used every customer in the control groups who viewed a product during the experimental period. 
The selection criteria meant that customers who saw a product and purchased it in the same browsing session did not qualify for the exposed groups, but still qualified for the control groups. These customers pushed up the metrics in the control groups, resulting in a negative incrementality in strategies.
In this case, we fixed the test by serving customers in the control groups charity adverts and only measuring customers who were served an ad.
The incident highlighted the need to include the individual incrementalities as a data quality metric when using the SIT framework.

\section{Conclusion}
\label{sec:conclusion}

We have presented the Stacked Incrementality Test (SIT) Framework for online controlled tests. The framework addresses the important problem of testing personalised user strategies where personalisation breaks statistical equivalence. We have provided a thorough analysis of the framework and provided a bound for when it is superior to the best simple A/B test. This bound is often met by the parameters of modern online controlled tests. As no power calculator exists that can be configured to the SIT framework, we have provided an open source power calculator for the benefit of the community.
In the final section, we focused on the pitfalls of running the SIT framework and lesson learnt via three case studies. While the incidents are specific to the framework, the learnings are general to online controlled experiments. We believe that sharing these lessons will allow practitioners to avoid our past mistakes.

%% file: sample-sigconf.bbl

\begin{thebibliography}{31}


\ifx \showCODEN    \undefined \def \showCODEN     #1{\unskip}     \fi
\ifx \showDOI      \undefined \def \showDOI       #1{#1}\fi
\ifx \showISBNx    \undefined \def \showISBNx     #1{\unskip}     \fi
\ifx \showISBNxiii \undefined \def \showISBNxiii  #1{\unskip}     \fi
\ifx \showISSN     \undefined \def \showISSN      #1{\unskip}     \fi
\ifx \showLCCN     \undefined \def \showLCCN      #1{\unskip}     \fi
\ifx \shownote     \undefined \def \shownote      #1{#1}          \fi
\ifx \showarticletitle \undefined \def \showarticletitle #1{#1}   \fi
\ifx \showURL      \undefined \def \showURL       {\relax}        \fi
\providecommand\bibfield[2]{#2}
\providecommand\bibinfo[2]{#2}
\providecommand\natexlab[1]{#1}
\providecommand\showeprint[2][]{arXiv:#2}

\bibitem[\protect\citeauthoryear{Backstrom and Kleinberg}{Backstrom and
  Kleinberg}{2011}]%
        {backstrom11network}
\bibfield{author}{\bibinfo{person}{Lars Backstrom} {and} \bibinfo{person}{Jon
  Kleinberg}.} \bibinfo{year}{2011}\natexlab{}.
\newblock \showarticletitle{Network Bucket Testing}. In
  \bibinfo{booktitle}{{\em WWW '11}}. \bibinfo{publisher}{ACM},
  \bibinfo{address}{New York, NY, USA}, \bibinfo{pages}{615--624}.
\newblock
\showISBNx{978-1-4503-0632-4}


\bibitem[\protect\citeauthoryear{Bakshy and Eckles}{Bakshy and Eckles}{2013}]%
        {bakshy13uncertainty}
\bibfield{author}{\bibinfo{person}{Eytan Bakshy} {and} \bibinfo{person}{Dean
  Eckles}.} \bibinfo{year}{2013}\natexlab{}.
\newblock \showarticletitle{Uncertainty in Online Experiments with Dependent
  Data: An Evaluation of Bootstrap Methods}. In \bibinfo{booktitle}{{\em KDD
  '13}}. \bibinfo{publisher}{ACM}, \bibinfo{address}{New York, NY, USA},
  \bibinfo{pages}{1303--1311}.
\newblock
\showISBNx{978-1-4503-2174-7}


\bibitem[\protect\citeauthoryear{Browne and Jones}{Browne and Jones}{2017}]%
        {browne17whatworks}
\bibfield{author}{\bibinfo{person}{Will Browne} {and} \bibinfo{person}{Mike~S.
  Jones}.} \bibinfo{year}{2017}\natexlab{}.
\newblock \bibinfo{title}{What Works In e-commerce - A Meta-analysis of 6700
  Online Experiments}.
\newblock
  \bibinfo{howpublished}{\url{https://www.qubit.com/wp-content/uploads/2017/12/qubit-research-meta-analysis.pdf}}.
    (\bibinfo{date}{Jun} \bibinfo{year}{2017}).
\newblock
\newblock
\shownote{White Paper.}


\bibitem[\protect\citeauthoryear{Chamberlain, Cardoso, Liu, Pagliari, and
  Deisenroth}{Chamberlain et~al\mbox{.}}{2017}]%
        {chamberlain17customer}
\bibfield{author}{\bibinfo{person}{Benjamin~Paul Chamberlain},
  \bibinfo{person}{\^{A}ngelo Cardoso}, \bibinfo{person}{C.H.~Bryan Liu},
  \bibinfo{person}{Roberto Pagliari}, {and} \bibinfo{person}{Marc~Peter
  Deisenroth}.} \bibinfo{year}{2017}\natexlab{}.
\newblock \showarticletitle{Customer Lifetime Value Prediction Using
  Embeddings}. In \bibinfo{booktitle}{{\em KDD '17}}. \bibinfo{publisher}{ACM},
  \bibinfo{address}{New York, NY, USA}, \bibinfo{pages}{1753--1762}.
\newblock
\showISBNx{978-1-4503-4887-4}


\bibitem[\protect\citeauthoryear{Crook, Frasca, Kohavi, and Longbotham}{Crook
  et~al\mbox{.}}{2009}]%
        {crook09sevenpitfalls}
\bibfield{author}{\bibinfo{person}{Thomas Crook}, \bibinfo{person}{Brian
  Frasca}, \bibinfo{person}{Ron Kohavi}, {and} \bibinfo{person}{Roger
  Longbotham}.} \bibinfo{year}{2009}\natexlab{}.
\newblock \showarticletitle{Seven Pitfalls to Avoid when Running Controlled
  Experiments on the Web}. In \bibinfo{booktitle}{{\em KDD '09}}.
  \bibinfo{publisher}{ACM}, \bibinfo{address}{New York, NY, USA},
  \bibinfo{pages}{1105--1114}.
\newblock
\showISBNx{978-1-60558-495-9}


\bibitem[\protect\citeauthoryear{Deng, Lu, and Litz}{Deng
  et~al\mbox{.}}{2017}]%
        {deng17trustworthy}
\bibfield{author}{\bibinfo{person}{Alex Deng}, \bibinfo{person}{Jiannan Lu},
  {and} \bibinfo{person}{Jonthan Litz}.} \bibinfo{year}{2017}\natexlab{}.
\newblock \showarticletitle{Trustworthy Analysis of Online A/B Tests: Pitfalls,
  Challenges and Solutions}. In \bibinfo{booktitle}{{\em WSDM '17}}.
  \bibinfo{publisher}{ACM}, \bibinfo{address}{New York, NY, USA},
  \bibinfo{pages}{641--649}.
\newblock
\showISBNx{978-1-4503-4675-7}


\bibitem[\protect\citeauthoryear{Deng and Shi}{Deng and Shi}{2016}]%
        {deng16datadriven}
\bibfield{author}{\bibinfo{person}{Alex Deng} {and} \bibinfo{person}{Xiaolin
  Shi}.} \bibinfo{year}{2016}\natexlab{}.
\newblock \showarticletitle{Data-Driven Metric Development for Online
  Controlled Experiments: Seven Lessons Learned}. In \bibinfo{booktitle}{{\em
  KDD '16}}. \bibinfo{publisher}{ACM}, \bibinfo{address}{New York, NY, USA},
  \bibinfo{pages}{77--86}.
\newblock
\showISBNx{978-1-4503-4232-2}


\bibitem[\protect\citeauthoryear{Deng, Xu, Kohavi, and Walker}{Deng
  et~al\mbox{.}}{2013}]%
        {deng13improving}
\bibfield{author}{\bibinfo{person}{Alex Deng}, \bibinfo{person}{Ya Xu},
  \bibinfo{person}{Ron Kohavi}, {and} \bibinfo{person}{Toby Walker}.}
  \bibinfo{year}{2013}\natexlab{}.
\newblock \showarticletitle{Improving the Sensitivity of Online Controlled
  Experiments by Utilizing Pre-experiment Data}. In \bibinfo{booktitle}{{\em
  WSDM '13}}. \bibinfo{publisher}{ACM}, \bibinfo{address}{New York, NY, USA},
  \bibinfo{pages}{123--132}.
\newblock
\showISBNx{978-1-4503-1869-3}


\bibitem[\protect\citeauthoryear{Dmitriev, Gupta, Kim, and Vaz}{Dmitriev
  et~al\mbox{.}}{2017}]%
        {dmitriev17adirtydozen}
\bibfield{author}{\bibinfo{person}{Pavel Dmitriev}, \bibinfo{person}{Somit
  Gupta}, \bibinfo{person}{Dong~Woo Kim}, {and} \bibinfo{person}{Garnet Vaz}.}
  \bibinfo{year}{2017}\natexlab{}.
\newblock \showarticletitle{A Dirty Dozen: Twelve Common Metric Interpretation
  Pitfalls in Online Controlled Experiments}. In \bibinfo{booktitle}{{\em KDD
  '17}}. \bibinfo{publisher}{ACM}, \bibinfo{address}{New York, NY, USA},
  \bibinfo{pages}{1427--1436}.
\newblock
\showISBNx{978-1-4503-4887-4}


\bibitem[\protect\citeauthoryear{Ellis}{Ellis}{2010}]%
        {ellis2010essential}
\bibfield{author}{\bibinfo{person}{P.D. Ellis}.}
  \bibinfo{year}{2010}\natexlab{}.
\newblock \bibinfo{booktitle}{{\em The Essential Guide to Effect Sizes:
  Statistical Power, Meta-Analysis, and the Interpretation of Research
  Results}}.
\newblock \bibinfo{publisher}{Cambridge University Press}.
\newblock
\showISBNx{9780521142465}
\showLCCN{2010007120}


\bibitem[\protect\citeauthoryear{Fleiss, Levin, and Paik}{Fleiss
  et~al\mbox{.}}{2004}]%
        {fleiss04determining}
\bibfield{author}{\bibinfo{person}{Joseph~L. Fleiss}, \bibinfo{person}{Bruce
  Levin}, {and} \bibinfo{person}{Myunghee~Cho Paik}.}
  \bibinfo{year}{2004}\natexlab{}.
\newblock \bibinfo{booktitle}{{\em Determining Sample Sizes Needed to Detect a
  Difference between Two Proportions}}.
\newblock \bibinfo{publisher}{John Wiley \& Sons, Inc.},
  \bibinfo{pages}{64--85}.
\newblock
\showISBNx{9780471445425}


\bibitem[\protect\citeauthoryear{Hill, Moakler, Hubbard, Tsemekhman, Provost,
  and Tsemekhman}{Hill et~al\mbox{.}}{2015}]%
        {hill15measuring}
\bibfield{author}{\bibinfo{person}{Daniel~N. Hill}, \bibinfo{person}{Robert
  Moakler}, \bibinfo{person}{Alan~E. Hubbard}, \bibinfo{person}{Vadim
  Tsemekhman}, \bibinfo{person}{Foster Provost}, {and} \bibinfo{person}{Kiril
  Tsemekhman}.} \bibinfo{year}{2015}\natexlab{}.
\newblock \showarticletitle{Measuring Causal Impact of Online Actions via
  Natural Experiments: Application to Display Advertising}. In
  \bibinfo{booktitle}{{\em KDD '15}}. \bibinfo{publisher}{ACM},
  \bibinfo{address}{New York, NY, USA}, \bibinfo{pages}{1839--1847}.
\newblock
\showISBNx{978-1-4503-3664-2}


\bibitem[\protect\citeauthoryear{Hill, Nassif, Liu, Iyer, and
  Vishwanathan}{Hill et~al\mbox{.}}{2017}]%
        {hill17efficient}
\bibfield{author}{\bibinfo{person}{Daniel~N. Hill}, \bibinfo{person}{Houssam
  Nassif}, \bibinfo{person}{Yi Liu}, \bibinfo{person}{Anand Iyer}, {and}
  \bibinfo{person}{S.V.N. Vishwanathan}.} \bibinfo{year}{2017}\natexlab{}.
\newblock \showarticletitle{An Efficient Bandit Algorithm for Realtime
  Multivariate Optimization}. In \bibinfo{booktitle}{{\em KDD '17}}.
  \bibinfo{publisher}{ACM}, \bibinfo{address}{New York, NY, USA},
  \bibinfo{pages}{1813--1821}.
\newblock
\showISBNx{978-1-4503-4887-4}


\bibitem[\protect\citeauthoryear{Hohnhold, O'Brien, and Tang}{Hohnhold
  et~al\mbox{.}}{2015}]%
        {hohnhold15focusing}
\bibfield{author}{\bibinfo{person}{Henning Hohnhold}, \bibinfo{person}{Deirdre
  O'Brien}, {and} \bibinfo{person}{Diane Tang}.}
  \bibinfo{year}{2015}\natexlab{}.
\newblock \showarticletitle{Focusing on the Long-term: It's Good for Users and
  Business}. In \bibinfo{booktitle}{{\em KDD '15}}. \bibinfo{publisher}{ACM},
  \bibinfo{address}{New York, NY, USA}, \bibinfo{pages}{1849--1858}.
\newblock
\showISBNx{978-1-4503-3664-2}


\bibitem[\protect\citeauthoryear{Johari, Koomen, Pekelis, and Walsh}{Johari
  et~al\mbox{.}}{2017}]%
        {johari17peeking}
\bibfield{author}{\bibinfo{person}{Ramesh Johari}, \bibinfo{person}{Pete
  Koomen}, \bibinfo{person}{Leonid Pekelis}, {and} \bibinfo{person}{David
  Walsh}.} \bibinfo{year}{2017}\natexlab{}.
\newblock \showarticletitle{Peeking at A/B Tests: Why It Matters, and What to
  Do About It}. In \bibinfo{booktitle}{{\em KDD '17}}.
  \bibinfo{publisher}{ACM}, \bibinfo{address}{New York, NY, USA},
  \bibinfo{pages}{1517--1525}.
\newblock
\showISBNx{978-1-4503-4887-4}


\bibitem[\protect\citeauthoryear{Kohavi, Deng, Frasca, Longbotham, Walker, and
  Xu}{Kohavi et~al\mbox{.}}{2012}]%
        {kohavi12trustworthy}
\bibfield{author}{\bibinfo{person}{Ron Kohavi}, \bibinfo{person}{Alex Deng},
  \bibinfo{person}{Brian Frasca}, \bibinfo{person}{Roger Longbotham},
  \bibinfo{person}{Toby Walker}, {and} \bibinfo{person}{Ya Xu}.}
  \bibinfo{year}{2012}\natexlab{}.
\newblock \showarticletitle{Trustworthy Online Controlled Experiments: Five
  Puzzling Outcomes Explained}. In \bibinfo{booktitle}{{\em KDD '12}}.
  \bibinfo{publisher}{ACM}, \bibinfo{address}{New York, NY, USA},
  \bibinfo{pages}{786--794}.
\newblock
\showISBNx{978-1-4503-1462-6}


\bibitem[\protect\citeauthoryear{Kohavi, Deng, Frasca, Walker, Xu, and
  Pohlmann}{Kohavi et~al\mbox{.}}{2013}]%
        {kohavi13online}
\bibfield{author}{\bibinfo{person}{Ron Kohavi}, \bibinfo{person}{Alex Deng},
  \bibinfo{person}{Brian Frasca}, \bibinfo{person}{Toby Walker},
  \bibinfo{person}{Ya Xu}, {and} \bibinfo{person}{Nils Pohlmann}.}
  \bibinfo{year}{2013}\natexlab{}.
\newblock \showarticletitle{Online Controlled Experiments at Large Scale}. In
  \bibinfo{booktitle}{{\em KDD '13}}. \bibinfo{publisher}{ACM},
  \bibinfo{address}{New York, NY, USA}, \bibinfo{pages}{1168--1176}.
\newblock
\showISBNx{978-1-4503-2174-7}


\bibitem[\protect\citeauthoryear{Kohavi, Deng, Longbotham, and Xu}{Kohavi
  et~al\mbox{.}}{2014}]%
        {kohavi14sevenrules}
\bibfield{author}{\bibinfo{person}{Ron Kohavi}, \bibinfo{person}{Alex Deng},
  \bibinfo{person}{Roger Longbotham}, {and} \bibinfo{person}{Ya Xu}.}
  \bibinfo{year}{2014}\natexlab{}.
\newblock \showarticletitle{Seven Rules of Thumb for Web Site Experimenters}.
  In \bibinfo{booktitle}{{\em KDD '14}}. \bibinfo{publisher}{ACM},
  \bibinfo{address}{New York, NY, USA}, \bibinfo{pages}{1857--1866}.
\newblock
\showISBNx{978-1-4503-2956-9}


\bibitem[\protect\citeauthoryear{Kohavi and Longbotham}{Kohavi and
  Longbotham}{2011}]%
        {kohavi11unexpected}
\bibfield{author}{\bibinfo{person}{Ron Kohavi} {and} \bibinfo{person}{Roger
  Longbotham}.} \bibinfo{year}{2011}\natexlab{}.
\newblock \showarticletitle{Unexpected Results in Online Controlled
  Experiments}.
\newblock \bibinfo{journal}{{\em SIGKDD Explor. Newsl.\/}}
  \bibinfo{volume}{12}, \bibinfo{number}{2} (\bibinfo{date}{March}
  \bibinfo{year}{2011}), \bibinfo{pages}{31--35}.
\newblock
\showISSN{1931-0145}


\bibitem[\protect\citeauthoryear{Kohavi, Longbotham, Sommerfield, and
  Henne}{Kohavi et~al\mbox{.}}{2009}]%
        {kohavi09controlled}
\bibfield{author}{\bibinfo{person}{Ron Kohavi}, \bibinfo{person}{Roger
  Longbotham}, \bibinfo{person}{Dan Sommerfield}, {and}
  \bibinfo{person}{Randal~M. Henne}.} \bibinfo{year}{2009}\natexlab{}.
\newblock \showarticletitle{Controlled experiments on the web: survey and
  practical guide}.
\newblock \bibinfo{journal}{{\em Data Mining and Knowledge Discovery\/}}
  \bibinfo{volume}{18}, \bibinfo{number}{1} (\bibinfo{date}{01 Feb}
  \bibinfo{year}{2009}), \bibinfo{pages}{140--181}.
\newblock
\showISSN{1573-756X}


\bibitem[\protect\citeauthoryear{Livingston and Cassidy}{Livingston and
  Cassidy}{2005}]%
        {livingston05statistical}
\bibfield{author}{\bibinfo{person}{Edward~H Livingston} {and}
  \bibinfo{person}{Laura Cassidy}.} \bibinfo{year}{2005}\natexlab{}.
\newblock \showarticletitle{Statistical power and estimation of the number of
  required subjects for a study based on the t-test: a surgeon's primer}.
\newblock \bibinfo{journal}{{\em The Journal of surgical research\/}}
  \bibinfo{volume}{126}, \bibinfo{number}{2} (\bibinfo{date}{June}
  \bibinfo{year}{2005}), \bibinfo{pages}{149--159}.
\newblock
\showISSN{0022-4804}


\bibitem[\protect\citeauthoryear{Lu and Liu}{Lu and Liu}{2014}]%
        {lu14separation}
\bibfield{author}{\bibinfo{person}{Luo Lu} {and} \bibinfo{person}{Chuang Liu}.}
  \bibinfo{year}{2014}\natexlab{}.
\newblock \bibinfo{title}{Separation strategies for three pitfalls in A/B
  testing}.  (\bibinfo{date}{Aug} \bibinfo{year}{2014}).
\newblock
\showURL{%
\url{http://www.ueo-workshop.com/wp-content/uploads/2014/04/Separation-strategies-for-three-pitfalls-in-AB-testing_withacknowledgments.pdf}}
\newblock
\shownote{Presented in The Second Workshop on User Engagement Optimization at
  KDD 2014, New York City, USA.}


\bibitem[\protect\citeauthoryear{Moss}{Moss}{2014}]%
        {moss14experiment}
\bibfield{author}{\bibinfo{person}{Will Moss}.}
  \bibinfo{year}{2014}\natexlab{}.
\newblock \bibinfo{title}{Experiment Reporting Framework}.
\newblock
  \bibinfo{howpublished}{\url{https://medium.com/airbnb-engineering/experiment-reporting-framework-4e3fcd29e6c0}}.
    (\bibinfo{date}{May} \bibinfo{year}{2014}).
\newblock
\newblock
\shownote{Blog post.}


\bibitem[\protect\citeauthoryear{Poyarkov, Drutsa, Khalyavin, Gusev, and
  Serdyukov}{Poyarkov et~al\mbox{.}}{2016}]%
        {poyarkov16boosted}
\bibfield{author}{\bibinfo{person}{Alexey Poyarkov}, \bibinfo{person}{Alexey
  Drutsa}, \bibinfo{person}{Andrey Khalyavin}, \bibinfo{person}{Gleb Gusev},
  {and} \bibinfo{person}{Pavel Serdyukov}.} \bibinfo{year}{2016}\natexlab{}.
\newblock \showarticletitle{Boosted Decision Tree Regression Adjustment for
  Variance Reduction in Online Controlled Experiments}. In
  \bibinfo{booktitle}{{\em KDD '16}}. \bibinfo{publisher}{ACM},
  \bibinfo{address}{New York, NY, USA}, \bibinfo{pages}{235--244}.
\newblock
\showISBNx{978-1-4503-4232-2}


\bibitem[\protect\citeauthoryear{Sadler}{Sadler}{2015}]%
        {sadler15whynot}
\bibfield{author}{\bibinfo{person}{George Sadler}.}
  \bibinfo{year}{2015}\natexlab{}.
\newblock \bibinfo{title}{Why Not Treat Marketing Like Sales?}
\newblock
  \bibinfo{howpublished}{\url{https://channels.theinnovationenterprise.com/presentations/why-not-treat-marketing-like-sales}}.
    (\bibinfo{date}{Jan} \bibinfo{year}{2015}).
\newblock
\newblock
\shownote{Presented in Business Analytics Innovation Summit (hosted by The
  Innovation Enterprise Ltd.), Las Vegas, USA.}


\bibitem[\protect\citeauthoryear{Satterthwaite}{Satterthwaite}{1946}]%
        {satterthwaite46}
\bibfield{author}{\bibinfo{person}{F.~E. Satterthwaite}.}
  \bibinfo{year}{1946}\natexlab{}.
\newblock \showarticletitle{An Approximate Distribution of Estimates of
  Variance Components}.
\newblock \bibinfo{journal}{{\em Biometrics Bulletin\/}} \bibinfo{volume}{2},
  \bibinfo{number}{6} (\bibinfo{year}{1946}), \bibinfo{pages}{110--114}.
\newblock
\showISSN{00994987}


\bibitem[\protect\citeauthoryear{Tang, Agarwal, O'Brien, and Meyer}{Tang
  et~al\mbox{.}}{2010}]%
        {tang10overlapping}
\bibfield{author}{\bibinfo{person}{Diane Tang}, \bibinfo{person}{Ashish
  Agarwal}, \bibinfo{person}{Deirdre O'Brien}, {and} \bibinfo{person}{Mike
  Meyer}.} \bibinfo{year}{2010}\natexlab{}.
\newblock \showarticletitle{Overlapping Experiment Infrastructure: More,
  Better, Faster Experimentation}. In \bibinfo{booktitle}{{\em KDD '10}}.
  \bibinfo{publisher}{ACM}, \bibinfo{address}{New York, NY, USA},
  \bibinfo{pages}{17--26}.
\newblock
\showISBNx{978-1-4503-0055-1}


\bibitem[\protect\citeauthoryear{Welch}{Welch}{1947}]%
        {welch47}
\bibfield{author}{\bibinfo{person}{B.~L. Welch}.}
  \bibinfo{year}{1947}\natexlab{}.
\newblock \showarticletitle{The Generalization of `Student's' Problem when
  Several Different Population Variances are Involved}.
\newblock \bibinfo{journal}{{\em Biometrika\/}} \bibinfo{volume}{34},
  \bibinfo{number}{1/2} (\bibinfo{year}{1947}), \bibinfo{pages}{28--35}.
\newblock
\showISSN{00063444}


\bibitem[\protect\citeauthoryear{Xie and Aurisset}{Xie and Aurisset}{2016}]%
        {xie16improving}
\bibfield{author}{\bibinfo{person}{Huizhi Xie} {and} \bibinfo{person}{Juliette
  Aurisset}.} \bibinfo{year}{2016}\natexlab{}.
\newblock \showarticletitle{Improving the Sensitivity of Online Controlled
  Experiments: Case Studies at Netflix}. In \bibinfo{booktitle}{{\em KDD '16}}.
  \bibinfo{publisher}{ACM}, \bibinfo{address}{New York, NY, USA},
  \bibinfo{pages}{645--654}.
\newblock
\showISBNx{978-1-4503-4232-2}


\bibitem[\protect\citeauthoryear{Xu and Chen}{Xu and Chen}{2016}]%
        {xu16evaluating}
\bibfield{author}{\bibinfo{person}{Ya Xu} {and} \bibinfo{person}{Nanyu Chen}.}
  \bibinfo{year}{2016}\natexlab{}.
\newblock \showarticletitle{Evaluating Mobile Apps with A/B and Quasi A/B
  Tests}. In \bibinfo{booktitle}{{\em KDD '16}}. \bibinfo{publisher}{ACM},
  \bibinfo{address}{New York, NY, USA}, \bibinfo{pages}{313--322}.
\newblock
\showISBNx{978-1-4503-4232-2}


\bibitem[\protect\citeauthoryear{Xu, Chen, Fernandez, Sinno, and Bhasin}{Xu
  et~al\mbox{.}}{2015}]%
        {xu15frominfrastructure}
\bibfield{author}{\bibinfo{person}{Ya Xu}, \bibinfo{person}{Nanyu Chen},
  \bibinfo{person}{Addrian Fernandez}, \bibinfo{person}{Omar Sinno}, {and}
  \bibinfo{person}{Anmol Bhasin}.} \bibinfo{year}{2015}\natexlab{}.
\newblock \showarticletitle{From Infrastructure to Culture: A/B Testing
  Challenges in Large Scale Social Networks}. In \bibinfo{booktitle}{{\em KDD
  '15}}. \bibinfo{publisher}{ACM}, \bibinfo{address}{New York, NY, USA},
  \bibinfo{pages}{2227--2236}.
\newblock
\showISBNx{978-1-4503-3664-2}


\end{thebibliography}
